\documentclass[prl,twocolumn,amsmath,amssymb,superscriptaddress]{revtex4-2}

\usepackage{graphicx}
\usepackage{bm}
\usepackage{color}
\usepackage{url}
\usepackage{tikz}
\usetikzlibrary{calc}
\usepackage{pgfplots}
\usepackage{amsmath,amssymb}
\usepackage{verbatim}
\usepackage{physics}
\usepackage{xcolor}
\usepackage{mwe}
\usepackage{tikz}
\usetikzlibrary{calc}
\usepackage{pgfplots}

\newcommand{\bea}{\begin{eqnarray}}
\newcommand{\eea}{\end{eqnarray}}

\newcommand{\be}{\begin{equation}}
\newcommand{\ee}{\end{equation}}

\newcommand{\beal}{\begin{align}}
\newcommand{\eeal}{\end{align}}

\newcommand{\ch}{\mathcal{H}}

\begin{document}


\title{Functionalization of $g$-wave altermagnets: spin-splitter effect enabled by surfaces}

\author{Sopheak Sorn}
\affiliation{Institute for Quantum Materials and Technologies, Karlsruhe Institute of Technology, 76131 Karlsruhe, Germany}
\affiliation{Institute of Theoretical Solid State Physics, Karlsruhe Institute of Technology, 76131 Karlsruhe, Germany}
\email{sopheak.sorn@kit.edu}

\begin{abstract}
We investigate surfaces of a $g$-wave altermagnet (AM) and show that they provide a platform for realizing $d$-wave altermagnetism and the associated spin-splitter functionality. Using the Kubo formalism applied to a minimal slab model, we evaluate the spin-splitter effect (SSE) by computing the spin conductivity corresponding to a transverse spin current induced by a longitudinal electric field. We find a finite SSE, absent in the bulk, that emerges from surface-induced $d$-wave altermagnetism. Strikingly, the sign pattern of the $d$-wave altermagnetism on both surfaces of the slab geometry is identical to each other, leading to \emph{additive} contributions to SSE from the two surfaces, with a spin-splitter angle reaching up to 15 degrees. In addition, this response is intrinsically linked to an accompanying surface-induced weak ferromagnetism, which potentially enables control of altermagnetic domains via an external magnetic field and provides a route to optimize the SSE functionality. These results can be understood in terms of a bulk–boundary correspondence between surface states and bulk altermagnetic order parameters, where the magnetic multipolar character of the latter plays a central role. Our findings strongly suggest thin-film engineering as a viable strategy to functionalize non-$d$-wave AMs.
\end{abstract}

\date{\today}

\maketitle

An emerging class of collinear compensated magnets, termed altermagnets (AMs), have recently attracted considerable attention owing to their characteristic spin-split electronic structure in spite of their vanishing magnetization \cite{LiborPRX, LiborPRX2, Bai2024, Hayami2019}. These systems are classified by the symmetry of their momentum-dependent spin splitting in the reciprocal space, denoted as $X$-wave ($X=d,g,i$), in close analogy to pairing symmetries in unconventional superconductors. From the spintronics application perspective, the most sought-after are $d$-wave AMs that are metallic thanks to their concomitant capability to convert a charge current into spin current without relying on the relativistic spin-orbit-coupling (SOC) strength\cite{Gonzalez2021}. This mechanism is known as the spin-splitter effect (SSE) and has been proposed as a promising route for efficiently generating spin currents in spintronic devices \cite{Gonzalez2021}. 

Despite intense research activities driven by both fundamental interest and technological potential, the realization of a metallic $d$-wave AM at ambient conditions remains rather illusive. In contrast, there has been more success in uncovering non-$d$-wave altermagetic materials, such as $g$-wave AM CrSb and MnTe, as evident in various photoemission spectroscopy studies\cite{Krempasky2023, Lee2024, Osumi2024, Reimers2024, Hariki2024, Ding2024, Zeng2024, Yang2025}. However, their non-$d$-wave character precludes a linear SSE response, restricting the effect to higher-order nonlinear regimes under external fields \cite{Ezawa2025} and thereby significantly reducing its efficiency. To overcome this limitation, recent proposals have focused on engineering $d$-wave altermagnetism via application of strain or external field on non-$d$-wave systems \cite{Chakraborty2024, Karetta2025, Guo2025}, by adsorption on antiferromagnetic surfaces \cite{Liu2025} , by exfoliation of antiferromagnets \cite{Mazin2023,Jana2025}, and by expanding the searching ground for altermagnetism in less direct settings, e.g., boundaries of antiferromagnets \cite{Yang2026, Leeb2026, Lange2026, Sasioglu2026}.

\begin{figure}[t]
    \centering
    \includegraphics[width=0.75\linewidth]{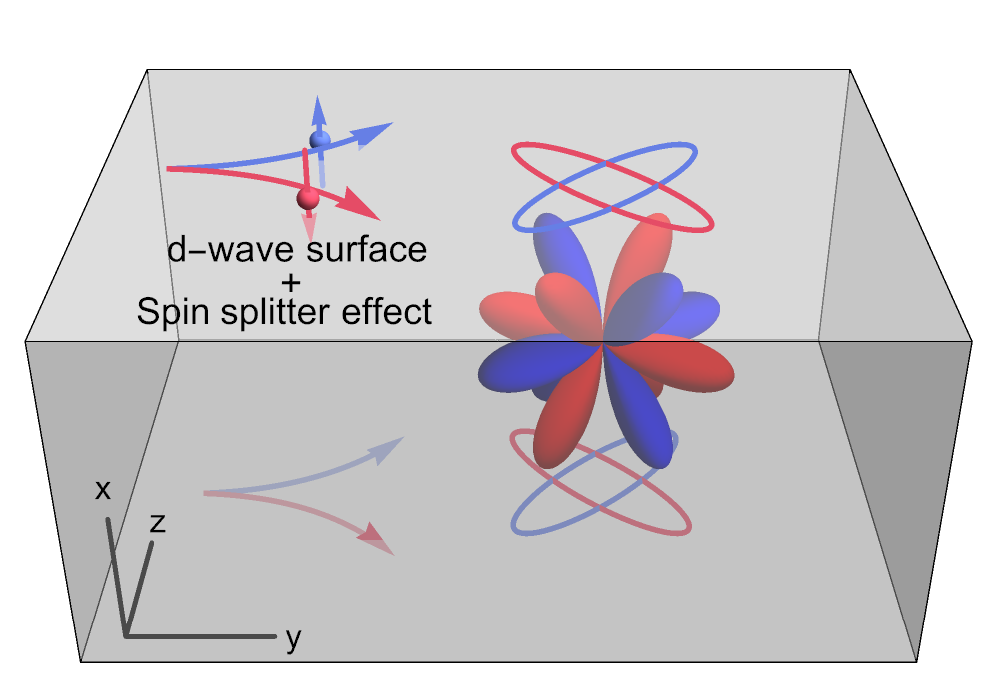}
    \caption{Schematic illustration of a bulk $g$-wave altermagnet in a slab geometry supporting surface $d$-wave spin splitting, thereby enabling the concomitant spin splitter effect---generation of transverse spin current upon applying a longitudinal charge current. This gives rise to the charge-to-spin conversion capability that is absent in the bulk.}
    \label{fig:fig1}
\end{figure}

\begin{figure*}[t]
    \centering
    \includegraphics[width=0.98\linewidth]{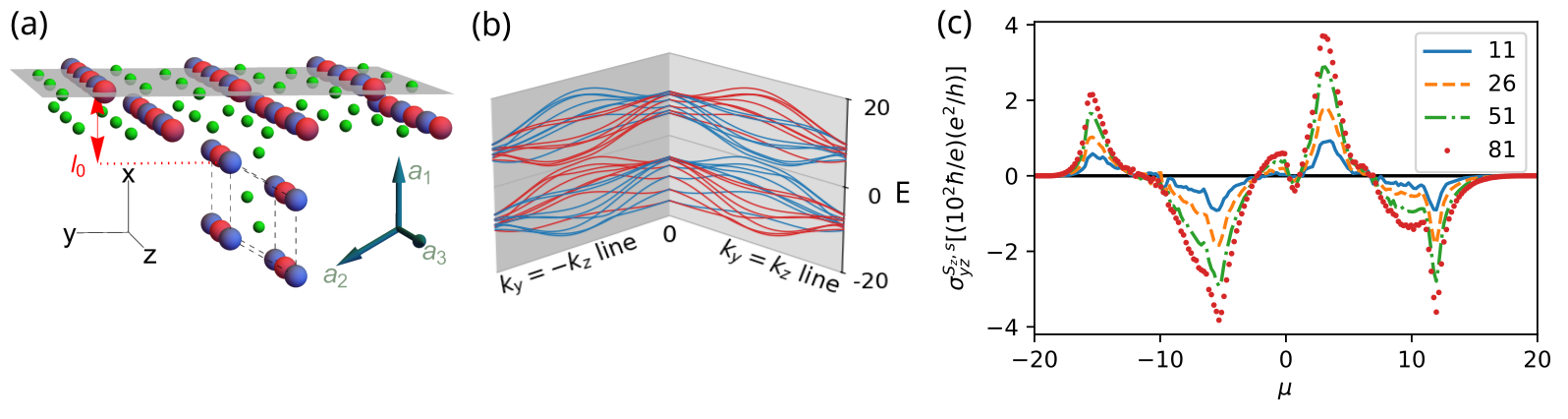}
    \caption{Panel (a) illustrates the crystal structure of the $(2\overline{1}0)$ surface of the $g$-wave AM model, which is inspired by the CrSb crystal. The box enclosed by the dashed lines represents the unit cell of the bulk. Blue and red spheres correspond to magnetic sublattices which host the N\'eel vector $\mathbf{N} \parallel \hat{z}$ in the altermagnetic state. Green spheres represent the nonmagnetic sites. The bulk lattice vectors are $\mathbf{a}_1 = a(1, 0, 0)^T$, $\mathbf{a}_2 = a(-1/2, \sqrt{3}/2, 0)^T$, and $\mathbf{a}_3 = c(0, 0, 1)^T$. Panel (b) displays the $d$-wave feature of the electronic spectrum obtained from slab-geometry diagonalization (results are for the film thickness $4l_0$.) Panel (c) displays the spin-splitter effect measured by the spin conductivity $\sigma^{S_x, s}_{yz}$ as a function of the chemical potential for slabs with different thicknesses: $11l_0$, $26l_0$, $51l_0$, and $81l_0$. Values of the model parameters for this plot are given in SMs.}
    \label{fig:fig2}
\end{figure*}

In this work, we theoretically demonstrate another viable route to $d$-wave altermagnetism and its SSE functionality: surfaces of a $g$-wave AM. Through the combination of a magnetic group analysis and $abinitio$ results in a companion paper \cite{Sorn2026}, the author and collaborators have formulated a bulk-boundary correspondence between the order parameter (OP) of a bulk $g$-wave AM and the $d$-wave character of the electronic band splitting at an optimal surface orientation of a slab geometry. Here, motivated by this, we scrutinize the accompanying SSE by applying Kubo formula to compute SSE \cite{Gonzalez2021} in a minimal $g$-wave altermagnetic model within a slab geometry. We demonstrate that a $g$-wave altermagnetic slab with an appropriate surface orientation indeed supports $d$-wave altermagnetism on both surfaces. Strikingly, the sign of the $d$-wave character is the \emph{same} on the two surfaces, leading to a constructive addition of their SSE contributions, as illustrated in Fig. \ref{fig:fig1}.  
We also observe three other remarkable features: (1) the efficiency of converting charge current into spin current, characterized by an angle that compares the SSE spin conductivity to the charge conductivity, can reach up to 15 degrees in our model. This value is comparable to that achieved in the strain-based proposal\cite{Karetta2025}. (2) Due to the low symmetry of the slab, there exist different principal axes with different values of the charge conductivity---a feature that can be harnessed to optimize the SSE angle. (3) The SSE induced by the surfaces is intertwined with a weak ferromagnetism, which may be exploited to manipulate altermagnetic domains via an external Zeeman magnetic field. Such weak ferromagnetism is absent in the bulk, consistent with a magnetic multipolar character of the altermagnetic OP, which forbids the usual Zeeman coupling.  
The surface-induced Zeeman coupling could be utilized to achieve altermagnetic mono-domain states in order to maximize the SSE. Our proposal has two attractive qualities: (i) parent metallic $g$-wave compounds are readily available, e.g., CrSb, so our work calls for experimental investigations into thin films thereof with the appropriate surface orientations. (ii) Conceptually, our proposal is based on a fundamental connection between the surface and bulk properties, which is governed by the magnetic multipolar nature of the altermagnetic OP \cite{Steward2023, Spaldin2024, McClarty2024, Schiff2024}.

\noindent \textcolor{blue}{\textit{Minimal model for $g$-wave altermagnets}---}To demonstrate the main results of our work, we employ a minimal single-orbital tight-binding model for $g$-wave AMs on a lattice that is isostructural to that of the $g$-wave AMs, CrSb and MnTe, as shown in Fig.\ref{fig:fig2}(a). The dashed lines enclose the unit cell of the bulk crystal, which consists of two magnetic sublattices, $A$ and $B$, denoted by red and blue spheres, respectively, whereas the green spheres represent nonmagnetic sites. The bulk Bloch Hamiltonian corresponding to electrons hopping between the magnetic sites is given by \cite{Roig2024}
\bea
    \ch &=& \ch_0 + \ch_{\rm soc},\\
    \ch_0 &=& t_{1, \mathbf{k}} + (t_{2, \mathbf{k}} + t_{3, \mathbf{k}}) \tau_x + t_{\text{AM}, \mathbf{k}} \tau_z - J \tau_z \mathbf{N} \cdot \boldsymbol{\sigma},\\
    \ch_{\rm soc} &=& \tau_y\boldsymbol{\lambda}_k \cdot \boldsymbol{\sigma},
\eea
where $t_{1,\mathbf{k}}$, $t_{2,\mathbf{k}}$, $t_{3,\mathbf{k}}$, $t_{\text{AM}, \mathbf{k}}$, and $\boldsymbol{\lambda}_{\mathbf{k}}$ are scalar functions whose forms are given in the Supplementary Materials (SM). $\boldsymbol{\tau}$ and $\boldsymbol{\sigma}$ are Pauli matrices acting on the sublattice space and spin space, respectively. The $J$ term is the exchange coupling with the N\'eel vector $\mathbf{N}$, which is set to be on the $z$-axis (along $\mathbf{a}_3$ lattice unit vector in Fig.\ref{fig:fig2}(a)), a feature inspired by CrSb. In the limit of vanishing SOC, $\ch_{\rm soc} = 0$, the 3D bulk band structure exhibits altermagnetic $S_z$-spin splitting with a momentum dependence given by a $g$-wave form factor $F(\mathbf{k})$ at small $|\mathbf{k}|$ \cite{Fernandes2024}: $F(\mathbf{k})=k_yk_z(3k_x^2 - k_y^2)$, whose $\pm$ sign in the 3D reciprocal space is represented by the blue and red lobes inside the slab in Fig.\ref{fig:fig1}. There are nodal planes, wherein the bands become spin degenerate, including the $k_xk_y$-plane and three other planes, namely $k_xk_z$-plane and its three-fold-rotation $C_{3z}$ images. Notably, the $g$-wave altermagnetic OP coincides with a rank-5 magnetic multipole moment $\varphi$ which can be expressed in terms of $F(\mathbf{k})$ as $\varphi \propto F(\mathbf{k})\mu_z$, where $\mu_z$ is a dipole moment in the $z$-direction \cite{Sorn2026, McClarty2024, Schiff2024, Fernandes2024}.

Using the real-space form of $\mathcal{H}$, we compute the electronic spectrum for a slab geometry with the $(2\overline{1}0)$ surface orientation \footnote{Here, I adopt the Miller index convention to label the surface orientation, and I have chosen the lattice unit vectors given in Fig.\ref{fig:fig2}(a).}, i.e., $yz$-plane, as shown in Fig.\ref{fig:fig2}(a). For $\ch_{\rm soc}$ set to zero, the bands are plotted in Fig.\ref{fig:fig2}(b) along the $k_y = \pm k_z$ lines within the 2D surface BZ corresponding to the $k_yk_z$-plane. The blue and red color denote electronic states with the spin polarization $\pm s_z$, respectively. Remarkably, we observe a $d$-wave spin splitting, which is clearly seen from the sign changing of the spin polarization upon a 90-degree rotation in Fig.\ref{fig:fig2}(b)(alongside with a spin-degenerate spectrum along $k_y$-axis and $k_z$-axis, which exists but is not shown explicitly here.)
This can be qualitatively characterized by a Zeeman-splitting term in the surface Bloch Hamiltonian featuring a momentum-dependent prefactor: $k_yk_z \sigma_z$ at small $|\mathbf{k}|$ \cite{Sorn2026}. 
Such a $d$-wave character naturally arises from how the bulk multipolar OP $\varphi$ manifests at the surface, as explained in Ref.\cite{Sorn2026}. There, we employed a group-theoretical method to show how the reduced symmetry at the $(2\overline{1}0)$ surface combines with bulk OP $\varphi$ to stabilize the surface $d$-wave spin splitting. 

Another perspective that provides more physical insights into this outcome is how the bulk multipole moments generally support boundary multipole moments of \emph{lower} ranks\cite{Vanderbilt1993,Benalcazar2017}. In our context, the rank-5 magnetic multipole OP $\varphi$ of the bulk $g$-wave AM stabilizes a rank-3 magnetic octupole moment at the $(2\overline{1}0)$ boundary, which is in turn responsible for the $d$-wave spin splitting \cite{Spaldin2024} near the surfaces. Such a phenomenon is analogous of the classic example of the accumulation of electric charge (charge monopole) at the boundary of a system that hosts a uniform bulk electric polarization (charge dipole)\cite{Vanderbilt1993}. Another analogous example is how the bulk electric multipole moments induce lower-rank boundary electric multipoles in higher-order topological insulators, which lays the foundation for the topological bulk-boundary correspondence therein\cite{Benalcazar2017} (in particular, see the supplementary material of Ref.\cite{Benalcazar2017} for a detailed derivation of surface multipoles from bulk multipoles.)

Last but not least, we also find that the two surfaces of the $(2\overline{1}0)$ slab geometry support the $d$-wave spin splitting \emph{with the same sign}. This has a profound consequence: both surfaces have the same-sign contribution to SSE, as illustrated by the bottom and top surface in Fig.\ref{fig:fig1}. This is studied next.

\noindent \textcolor{blue}{\textit{Spin-splitter effect}--} SSE measures the ability of a system to generate a transverse spin current upon application of a longitudinal electric field\cite{Gonzalez2021}. It differs from the spin-Hall counterpart by existing independently of the SOC strength and being odd under time-reversal operation \cite{Gonzalez2021}. Among all AMs, only the $d$-wave AMs support SSE \cite{Ezawa2025}, which is characterized by the symmetric part of the off-diagonal spin conductivity tensor, $\sigma_{yz}^{S_z, s} = (\sigma_{yz}^{S_z}+\sigma_{zy}^{S_z})/2$ \cite{Gonzalez2021}.
It describes how much spin $S_z$ current is generated along the $y$-direction ($z$-direction) in response to an electric field along the $z$-axis ($y$-axis).
We compute SSE by computing the spin conductivity $\sigma^{S_z}_{ab}$ using Kubo formula for the slab geometry; see Ref. \cite{Sorn2021, Sorn2025} and SM for details. Within the constant relaxation time $\tau$ approximation characterized by a small energy broadening $\gamma = \hbar/\tau$, SSE spin conductivity scales linearly with $1/\gamma$ \cite{Gonzalez2021}, which is analogous to the Drude form of the diagonal components of the charge conductivity tensor. In our computation, we choose a small $\gamma = 0.05$ in the energy unit used in Fig.\ref{fig:fig2}(b), thus much smaller than the band width.

Figure \ref{fig:fig2}(c) shows SSE $\sigma_{yz}^{S_z, s}$ as a function of the chemical potential $\mu$ for $(2\overline{1}0)$ slabs with different thicknesses, including $11 l_0, 26 l_0, 51l_0$ and $81l_0$, where $l_0$ is defined in Fig.\ref{fig:fig2}(a). Indeed, our results confirm the existence of SSE that accompanies the surface-induced $d$-wave character. We observe that $\sigma^{S_z, s}_{yz}$ increases upon increasing the thickness of the slab. However, to meaningfully examine the charge-to-spin conversion efficiency, we compare the SSE with the longitudinal charge conductivity by calculating the following SSE angle \cite{Gonzalez2021, Karetta2025}
\bea
    \theta_{aa} &=& \arctan\left(\frac{(2e/\hbar) \sigma^{S_z, s}_{yz}}{\sigma_{aa}}\right),
\eea
where $\sigma_{aa}$ is the longitudinal charge conductivity along $a = y,z$. The charge conductivity tensor features two principle axes along $a = y, z$, so that $\sigma_{zz}$ differs from $\sigma_{yy}$. This implies that the SSE angle $\theta_{aa}$ can be further maximized by choosing an optimal direction to apply the charge current. We note that due a similar scaling behavior with $1/\gamma$ in the small-$\gamma$ limit between $\sigma_{yz}^{S_z, a}$ and $\sigma_{aa}$, the SSE angle depends only weakly on $\gamma$\cite{Gonzalez2021}.

\begin{figure}[t]
    \centering
    \includegraphics[width=0.98\linewidth]{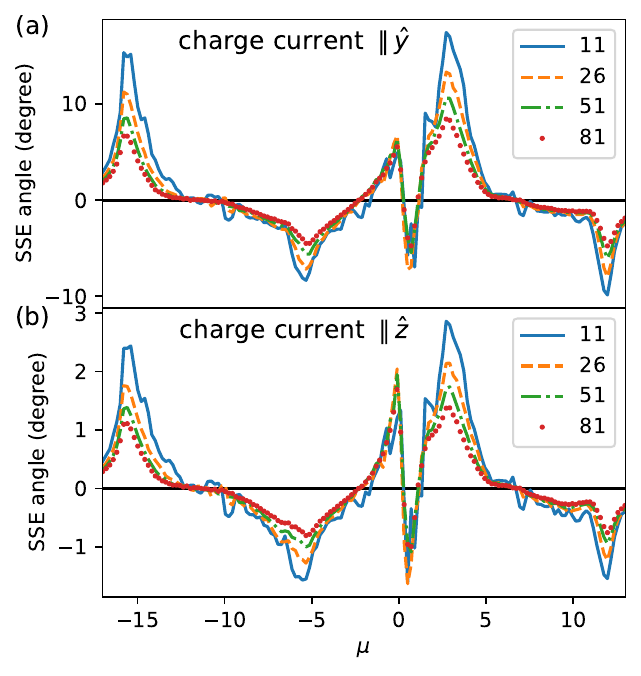}
    \caption{Spin-splitter angle as a function of the chemical potential $\mu$ for $(2\overline{1}0)$-oriented slab geometry with different thicknesses, including $11l_0, 26l_0, 51l_0,$ and $81l_0$, characterizing (a) the efficiency in generating a transverse $S_z$-spin current along z-axis upon applying charge current along y-axis and (b) the efficiency in generating a transverse $S_z$-spin current along y-axis upon applying charge current along z-axis.}
    \label{fig:angle}
\end{figure}

Figure \ref{fig:angle} shows the SSE angle in degree unit for the charge current along $y$-axis and $z$-axis in panel $(a)$ and $(b)$, respectively (they are resulted from applying electric field along $y$ and $z$ axis, respectively). Indeed, applying electric field along the $y$ principal axis generally results in a larger SSE angle than applying current along the $z$-axis. Remarkably, the SSE angle can become as large as $15$ degree; see panel (a). Moreover, we observe that the SSE angle in both panels decreases with the slab thickness, consistent with the fact that the SSE is induced by the surfaces. Although $\sigma_{yz}^{S_z, s}$ increases with the thickness, the charge conductivity increases at an even faster rate with the slab thickness, giving rise to the thickness dependence of the SSE angle in Fig.\ref{fig:angle}. 

These favorable results correspond to the slab geometry in a single-domain state. Since SSE is time-reversal odd, SSE contributions from opposite altermagnetic domains, i.e., $\mathbf{N} = \pm \hat{z}$, come with the opposite signs, leading to a cancellation effect when the system is in a random domain configuration. Due to the magnetic multipolar nature, altermagnetic domain manipulation can be challenging. For instance, the rank-5 nature of the bulk OP $\varphi$ in our model allows for a coupling to an external Zeeman magnetic field only beginning at the fifth order. This suggests a weak coupling, which, at the practical level, might not be harnessed to polarize the system into a single-domain state. In the remaining part of the manuscript, we discuss how the surfaces favorably induce weak ferromagnetism, which allows for a linear Zeeman coupling to an external magnetic field. This opens up the possibility of achieving a monodomain state by applied field or field cooling.

\begin{figure}[t]
    \centering
    \includegraphics[width=0.95\linewidth]{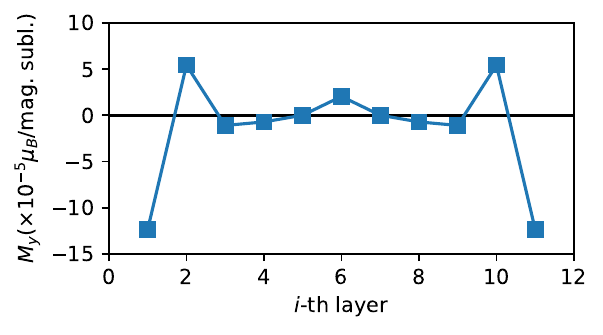}
    \caption{Layer-resolved $y$-component of the spin magnetization associated with the weak ferromagnetism in the slab with thickness $11l_0$. Here, we have chosen a particular value of the chemical potential $\mu$, as specified in SMs.}
    \label{fig:weakFM}
\end{figure}

\noindent \textcolor{blue}{\textit{Weak ferromagnetism}--}As the name suggests, this phenomenon scales with the small magnitude of the relativistic SOC effect and exists due to a sufficiently low magnetic symmetry (at the surfaces, in our case) that cannot constrain the magnetization to zero \cite{Dzialoshinskii1957}; see also Ref. \cite{Weber2024}. In the companion paper \cite{Sorn2026}, the author and collaborators have used a group theoretical method to show the existence of a weak magnetization in the y-direction, as permitted by the magnetic symmetry of the slab in the altermagnetic phase. Here, we compute the spin contribution to the magnetization, $M_y$, in our model. 
Figure \ref{fig:weakFM} shows the profile of $M_y$ per magnetic sublattice in different layers, where a layer is defined to be in the $(2\overline{1}0)$ plane with the thickness $l_0$ as shown in Fig.\ref{fig:fig1}(a). The results are straightforwardly obtained by computing the expectation values of Pauli spin operators associated with each layer. Note that we have set the $\ch_{\rm soc}$ term to be nonzero with a small coefficient to obtain the weak ferromagnetism (see SM). Clearly, the summation over the whole slab yields a nonzero magnetization value that is of the order of $10^{-5}\mu_B$, and this is a value that could potentially be resolved experimentally; see, for instance, Ref.\cite{Takuya2024} for a measurement of weak magnetization in MnTe at this order of magnitude. Consistent with its origin stemming from the surfaces, the strength of the ferromagnetism is strongest near the surfaces, corresponding to the first layer and the eleventh layer in Fig.\ref{fig:weakFM}, while the bulk region features a much smaller magnitude.
It is this weak magnetization in the $y$-direction that couples linearly to an external Zeeman magnetic field, thereby strongly suggesting the possibility to manipulate altermagnetic domains via an external field, as mentioned before.

\noindent \textcolor{blue}{\textit{Conclusion}---}Our work demonstrates, using a minimal $g$-wave AM model, that spin-splitter functionality, absent in bulk $g$-wave AMs, can be induced by surface effects. The existence of the surface-enabled SSE can be understood from a fundamental property of AMs that is valid beyond our minimal-model consideration: the magnetic multipolar OP of the $g$-wave AM imprints a $d$-wave character on surfaces, which is in turn responsible for the SSE.
We obtain a sizable spin-splitter angle of up to 15 degrees. Moreover, this functionality is intrinsically linked to a surface-induced weak ferromagnetism, which enables a linear Zeeman coupling to an external magnetic field. This coupling provides a potential route to control altermagnetic domains and thereby optimize the spin-splitter response. These findings point to thin-film engineering as a practical pathway for functionalizing non-$d$-wave AMs. 
Experimental verification of our predictions, e.g., in $g$-wave CrSb thin films, would be highly desirable.

\noindent \textcolor{blue}{\textit{Acknowledgment}--}The author would like to thank Markus Garst and Iksu Jang for very helpful discussions. The author also thanks Charanpreet Singh, Lukasz Plucinski, Gustav Bihlmayer, Yuriy Mokrousov, and Wulf Wulfhekel for many stimulating
discussions in a related project. The author is supported by the Deutsche Forschungsgemeinschaft through TRR 288 Grant No. 422213477 (Project No. A11).

\bibliographystyle{unsrtnat}
\bibliography{refs}

\end{document}



\title{Supplementary Material: Functionalization of $g$-wave altermagnets: spin-splitter effect enabled by surfaces}

\author{Sopheak Sorn}
\affiliation{Institute for Quantum Materials and Technologies, Karlsruhe Institute of Technology, 76131 Karlsruhe, Germany}
\affiliation{Institute of Theoretical Solid State Physics, Karlsruhe Institute of Technology, 76131 Karlsruhe, Germany}
\email{sopheak.sorn@kit.edu}

\pacs{}
\date{\today}

\maketitle

\beginsupplement
\section{Minimal model for $g$-wave altermagnets}

\begin{figure}[t]
    \centering
    \includegraphics[width=0.8\linewidth]{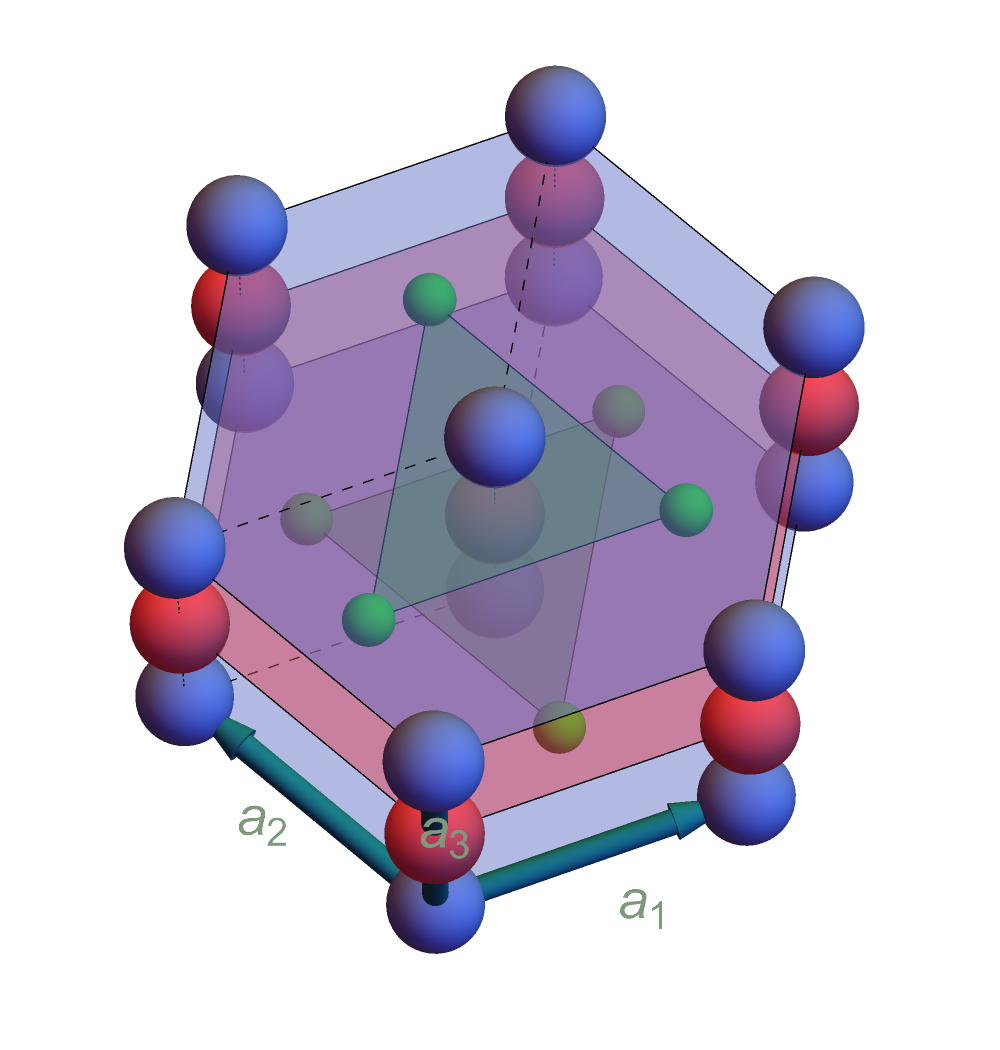}
    \caption{Illustration of the crystal structure of the minimal model. Red and blue sites correspond to magnetic sublattices, while the green spheres represent the nonmagnetic sites.}
    \label{fig:figS1}
\end{figure}

The minimal model is defined on a lattice that is isostructural to that of the $g$-wave altermagnet(AM), CrSb and MnTe. Figure \ref{fig:figS1} illustrates the crystal structure featuring a unit cell that consists of two magnetic sublattices, represented by blue and red spheres. Green spheres are the nonmagnetic sites which cause a crystal environment for the two magnetic sublattices to differ from each other. Due to the nonmagnetic sites, the two magnetic sublattices cannot be mapped into one another by an inversion--a characteristic feature of altermagnetism. The are related instead by a 3-fold rotation followed by a partial translation along the vertical direction. The lattice vectors are defined below
\bea
    \mathbf{a}_1 &=& a\left(1, 0, 0\right)^T,\\
    \mathbf{a}_2 &=& a\left(-\frac{1}{2}, \frac{\sqrt{3}}{2}, 0\right)^T,\\
    \mathbf{a}_3 &=& c\left(0, 0, 1\right)^T,
\eea
where $a$ and $c$ are variables. We set $a$ to unity and set $c=3a$ in our work. Within the unit cell whose position is at the origin, the two magnetic sublattices sit at: $(0,0,0)^T$ and $(0, 0, c/2)$. Meanwhile, the two nonmagnetic sites are at $(a/2, a\sqrt{3}/6, c/4)$ and $(a, a\sqrt{3}/3, 3c/4)$.  

\begin{widetext}

The coefficients in the Bloch Hamiltonian Eq.(1) in the main text are defined by \cite{Roig2024}
\bea
    t_{1,\bk} &=& -2t_1 \sum_{\mathbf{\boldsymbol{\delta}} = \boldsymbol{\delta}_{1N} } \cos\left(\mathbf{k}\cdot \boldsymbol{\delta}\right),\\
    t_{2, \bk} &=& 2t_2 \cos(\mathbf{k}\cdot \mathbf{a}_3),\\
    t_{3, \bk} &=& 2 t_3 \sum_{\boldsymbol{\delta} = \boldsymbol{\delta}_{3N}} \cos\left(\mathbf{k}\cdot \boldsymbol{\delta}\right),\\
    t_{\text{AM},\mathbf{k}} &=& 8t_{\rm AM} \sin(k_z c) \sum_{\boldsymbol{\delta} = \boldsymbol{\Delta}} \sin(\mathbf{k}\cdot \boldsymbol{\delta})
\eea
where $\boldsymbol{\delta}_{1N} \in \{\mathbf{a}_1, \mathbf{a}_2, -\mathbf{a}_1-\mathbf{a}_2\}$, $\boldsymbol{\delta}_{3N} = \boldsymbol{\delta}_{1N} + \mathbf{a}_3/2$, $\boldsymbol{\Delta} \in \{2\mathbf{a}_1 + \mathbf{a}_2, \mathbf{a}_2-\mathbf{a}_1, - 2\mathbf{a}_2 - \mathbf{a}_1\}$. The spin-orbit-coupling(SOC) terms are

\bea
    \lambda_{\bk, x} &=& 2 (c/a) \lambda \cos(k_z c/2) \left[\cos k_xa - \cos(k_xa/2)\cos(k_ya\sqrt{3}/2)\right],\nonumber\\
    \lambda_{\bk, y} &=& 2 (c/a) \sqrt{3} \lambda \cos(k_z c/2) \sin(k_x a/2) \sin(k_ya\sqrt{3}/2),\nonumber\\
    \lambda_{\bk, x} &=& 8 \lambda \sin(k_z c/2) \sin(k_x a/2) \left[\cos (k_xa/2) - \cos(k_ya\sqrt{3}/2)\right].\nonumber
\eea

These originate from the SOC hopping between A and B sublattices along the third-nearest-neighbor bonds defined by $\boldsymbol{\delta}_{3N}$. In the main text, we have set $\left(t_1, t_2, t_3, t_{\rm AM}, J, \lambda \right) = \left(1, 1, 1, 0.5, 8, 0.05\right)$. The chemical potential chosen for the weak ferromagnetism computation in Fig.4 in the main text is $\mu = -5t_1$.

\section{Kubo formula for the charge and spin conductivity}
In the main text, we show results obtained from applying the Kubo formula for charge $\sigma_{ab}$ and spin conductivity $\sigma^{S_z}_{ab}$ as given below; see also Ref.\cite{Sorn2021, Sorn2025}.

\bea
    \sigma_{ab} &=& \lim_{\omega \rightarrow 0} \frac{i2\pi}{V} \frac{e^2}{h} \sum_{\bk, m, n} \frac{f(E_{\bk m}) - f(E_{\bk n})}{E_{\bk n} - E_{\bk m}} \left[\frac{(\hat{j}_a)^{mn}_{\bk}(\hat j_b)^{nm}_{\bk}}{\hbar \omega + i\gamma + E_{\bk m} - E_{\bk n}}\right],\\
    \sigma^{S_z}_{ab} &=& \lim_{\omega \rightarrow 0} \frac{i2\pi}{V} \frac{e^2}{h} \sum_{\bk, m, n} \frac{f(E_{\bk m}) - f(E_{\bk n})}{E_{\bk n} - E_{\bk m}} \left[\frac{(\hat{j}^{S_z}_a)^{mn}_{\bk}(\hat j_b)^{nm}_{\bk}}{\hbar \omega + i\gamma + E_{\bk m} - E_{\bk n}}\right],
\eea
\end{widetext}
where $V$ is the volume (or area in 2D); $f$ is the Fermi-Dirac distribution function; $\bk = (k_y, k_z)^T$ in the $(2\overline{1}0)$ slab in our case; $E_{\bk m}$ is the energy eigenvalue; $m$ and $n$ are band indices; the matrix elements of the (spin) current operator $(\hat{j}_a)^{mn}_{\bk}$ ($(\hat{j}^{S_z}_a)^{mn}_{\bk}$) are defined at the end; $\gamma$ is a small broadening arising from a phenomenological lifetime of the electron, e.g. the transport lifetime. We have fixed a small $\gamma = 0.05t_1$ throughout our work. The current operator $\hat j$ is obtained from Peierls substitution in each hopping term in the tight-binding Hamiltonian, $t_{il, jl'}c^{\dagger}_{il}c_{jl'}$, between site $i$ and $j$ with the compact notation for spin and sublattice denoted by $l$ and $l'$:
\bea
    t_{il, jl'}c^{\dagger}_{il}c_{jl'} &\rightarrow& t_{il, jl'}c^{\dagger}_{il}c_{jl'} \exp\left(i \int_i^j d\mathbf{r}\cdot \mathbf{A}\right),\\
    &\approx & t_{il, jl'}c^{\dagger}_{il}c_{jl'} \left(1 + i \mathbf{r}_{ij}\frac{\mathbf{A}(i) + \mathbf{A}(j)}{2}\right),
\eea
where $\mathbf{r}_{ij} = \mathbf{r}_j - \mathbf{r}_i$, and $\mathbf{A}$ is a vector potential corresponding to an externally applied electric field $\mathbf{E} = -\partial \mathbf{A}/\partial t$. The current operator at a site $i$ is given by
\bea
    \hat{\mathbf{j}}(i) &=& \frac{\delta H[\mathbf{A}]}{\delta \mathbf{A}(i)},
\eea
where $H[\mathbf{A}]$ is the Hamiltonian after the Peierls substitution. The current operator in the Kubo formula is defined by
\bea
    \hat{j}_a &=& \sum_i \hat{j}_a(i).
\eea
Owing to the translational invariance, we can express the current operator as (see Appendix B of Ref. \cite{Sorn2025})
\bea
    \hat{j}_a &=& \sum_{\bk, ll'} c^{\dagger}_{\bk l} \mathcal{J}^a_{ll'} (\bk) c_{\bk l'},
\eea\
where $\mathcal{J}^a_{ll'}(\bk) = - \frac{e}{\hbar} \frac{\partial \ch_{ll'}(\bk)}{\partial k_a}$. 

The spin operator in the Kubo formula takes a similar form
\bea
    \hat{j}^{S_z}_a &=& \sum_{\bk ll'} c^{\dagger}_{\bk l} \mathcal{J}^{S_z, a}_{ll'}(\bk) c_{\bk l'},
\eea
where $\mathcal{J}^{S_z, a}_{ll'}(\bk) = \frac{1}{4}\{\sigma^z, \mathcal{J}^a_{ll'}(\bk)\}$. Finally, the matrix elements of the charge and spin current operators, appearing in the Kubo formulae, are
\bea
    (\hat{j}_a)^{mn}_{\bk} &=&  \bra{\bk m} \mathcal{J}^a(\mathbf{k})\ket{\bk n},\\
    (\hat{j}^{S_z}_a)^{mn}_{\bk} &=&  \bra{\bk m} \mathcal{J}^{S_z, a}(\mathbf{k})\ket{\bk n},
\eea
where $\ket{\bk n}$ is the eigenvector of the Bloch Hamiltonian $\ch(\bk)$.

\bibliographystyle{unsrtnat}
\bibliography{refs}